\documentclass[doublecol]{epl2}

\usepackage[english]{babel}
\usepackage{graphicx}
\usepackage{epsfig}
\usepackage{amssymb}
\usepackage{dcolumn}
\usepackage{bm}
\usepackage{color}
\usepackage{float}
\usepackage{hyperref} 
\hypersetup{colorlinks=true}

\newcommand{\BFCA}{\ensuremath{\mathrm{Ba(Fe}_{1-x}\mathrm{Co}_{x})_{2}\mathrm{As}_{2}}}

\newcommand{\BFCMA}{\ensuremath{\mathrm{Ba(Fe}_{1-x-y}\mathrm{Co}_{x}\mathrm{Mn}_{y})_{2}\mathrm{As}_{2}}}
\newcommand{\BFAP}{\ensuremath{\mathrm{BaFe}_{2}\mathrm{As}_{2-x}\mathrm{P}_{x}}}
\newcommand{\EFAP}{\ensuremath{\mathrm{EuFe}_{2}\mathrm{As}_{2-x}\mathrm{P}_{x}}}

\newcommand{\BFA}{\ensuremath{\mathrm{BaFe}_{2}\mathrm{As}_{2}}}
\newcommand{\KBFA}{\ensuremath{\mathrm{K}_{x}\mathrm{Ba}_{1-x}\mathrm{Fe}_{2}\mathrm{As}_{2}}}
\newcommand{\KEFA}{\ensuremath{\mathrm{K}_{x}\mathrm{Eu}_{1-x}\mathrm{Fe}_{2}\mathrm{As}_{2}}}
\newcommand{\dg}{\ensuremath{^\circ}}

\title{Electronic structure and quantum criticality in Ba(Fe$_{1-x-y}$Co$_{x}$Mn$_{y}$)$_{2}$As$_{2}$, an ARPES study.}

\author{E. D. L.  Rienks\inst{1} \and T. Wolf\inst{2} \and  K. Koepernik\inst{3} \and I. Avigo\inst{4} \and P. Hlawenka\inst{1} \and C. Lupulescu\inst{5} \and T. Arion\inst{6} \and F. Roth\inst{3,7} \and W. Eberhardt\inst{5,7} \and U. Bovensiepen\inst{4} \and J. Fink\inst{3}} \shortauthor{E.\,D.\,L. Rienks et\,al.}

\institute{
\inst{1} Helmholtz-Zentrum Berlin, Albert-Einstein-Strasse 15, 12489 Berlin, Germany\\
\inst{2} Karlsruhe Institute of Technology, Institut f\"ur Festk\"orperphysik, D-76021 Karlsruhe, Germany\\
\inst{3} Leibniz-Institute for Solid State and Materials Research Dresden, P.O.Box 270116, D-01171
Dresden, Germany\\
\inst{4} Fakult\"at f\"ur Physik, Universit\"at Duisburg-–Essen, Lotharstr. 1, D-47048 Duisburg, Germany\\
\inst{5} Technische Universit\"at Berlin, Institut f\"ur Optik und Atomare Physik, Strasse des 17. Juni 136, D-10623 Berlin, Germany\\
\inst{6} Institut f\"ur Experimentalphysik, Universit\"at Hamburg, Luruper Chaussee 149, 22761 Hamburg, Germany\\
\inst{7} Center for Free-Electron Laser Science / DESY, Notkestrasse 85, 22607 Hamburg, Germany\\
}

\date{\today}

\abstract{
We used angle-resolved photoemission spectroscopy (ARPES)  and density functional theory calculations to study the electronic structure  of \BFCMA\ for $x$=0.06 and $0\le y \le 0.07$. From ARPES we  derive that the  substitution of Fe by Mn does not lead to hole doping,  indicating a localization of the induced holes.
An  evaluation of the measured spectral function does not indicate a diverging effective mass or scattering rate near optimal doping. Thus the present ARPES results indicate a continuous  evolution of the quasiparticle interaction and therefore question previous quantum critical scenarios.
}


\begin{document}

\maketitle

In many compounds such as heavy fermion systems, doped cuprates, molecular crystals, and ferropnictides\,\cite{Johnston2010,Stewart2011}, unconventional/high--temperature  superconductivity occurs close to a point in the phase diagram where  at zero temperature the  antiferromangetic order vanishes. The transition into the superconducting region can be induced  by pressure, chemical pressure, doping, or some other control parameter.  Different from a temperature-driven transition, large quantum fluctuations are expected in the zero temperature phase transitions which led to the introduction of  the term quantum criticality\,\cite{Loehneysen2007,Gegenwart2008}. A  non-Fermi-liquid behavior is detected near the quantum critical point, which is related to a diverging effective mass and scattering rate and to a disappearance of the quasiparticle spectral weight at the Fermi level. The general explanation for the deviation from a Fermi liquid behavior in this region of the phase diagram is that electrons strongly interact with collective modes related to the antiferromagnetic (AFM) order. It is supposed that the superconducting phase, appearing close to the quantum critical point, is due to a coupling of the charge carriers to these quantum fluctuations. This would  provide a universal  explanation of the pairing mechanism in unconventional superconductors. Both quantum critical phenomena and unconventional superconductivity are major themes in current condensed matter physics.

For the iron pnictides the quantum criticality scenario has been proposed theoretically by Dai et al.\,\cite{Dai2009}.
Experimentally,  in ferropnictides, the scheme of a quantum critical point  is supported by transport and thermodynamic measurements on chemically pressurized  \BFAP\  or \EFAP\ \,\cite{Shishido2010,Kasahara2010,Hashimoto2012,Maiwald2012,Shibauchi2013}, on p-type doped \KBFA\ or \KEFA\ \,\cite{Gooch2009,Maiwald2012}, and n-type doped \BFCA\,\cite{Ning2010,Meingast2012}.

Angle-resolved photoemission spectroscopy (ARPES)\,\cite{Huefner2003}  is particularly suited to address the question whether a quantum critical scenario applies, since it  not only reveals the  Fermi surface topology, but offers also the unique possibility to determine the momentum dependent spectral function which contains information on the complex  self-energy $\Sigma=\Sigma$'+$i\Sigma$", which is related to the mass renormalization  and the scattering rate. Moreover we emphasize  that with momentum dependent ARPES data, in contrast to results from macroscopic methods, it is possible to distinguish between the hot and cold spots on the Fermi surface. In the ferropnictides, the hot spots are caused by a partial nesting between hole and electron pockets, yielding the AFM order,  potential AFM quantum fluctuations, and possibly also superconductivity. The cold spots  determine  the transport and thermodynamic properties in the normal state.  A variation  of the charge carrier dynamics at different points of the Fermi surface is expected due to the changes of the nesting conditions and/or the orbital character of the Fermi surface in $k$-space\,\cite{Hirschfeld2011}.

In this Letter we report an  electronic structure study of  \BFCMA\  (BFCMA) by means of ARPES. We focus on a series of crystals with $x$=0.06 and  a variable Mn substitution with $0\le y \le 0.07$. For $y=0$,  $T_c$  is close to its optimal value and close to the extrapolated zero temperature value of $T_N$. Thus within a quantum critical scenario we should find diverging effective masses and  scattering rates. With increasing Mn substitution previous studies on the same system have detected  a rapid decrease of $T_c$, a disappearance of superconductivity near $y=0.02$\,\cite{Li2012}, and a $T_N$ which  reaches a maximum of 23 K at $y$=0.01 and disappears near $y$=0.03\,\cite{Hardy2009}. This  signals that with increasing Mn substitution one departs from a possible quantum critical point and therefore the scattering rates and the mass renormalization should be reduced.

 In the most simple picture,  the substitution of Fe by Mn leads to hole doping and thus one should move to a lower total electron number. However, studies on Mn substituted ferropnictides  indicate that the case is by far more complex. Although  $T_N$  decreases with increasing Mn concentration, a superconducting dome does not appear\,\cite{Thaler2011}. Rather a transition into a new magnetic phase at a Mn concentration near 12 \% is discussed\,\cite{Inosov2013}.  This is in stark contrast to the hole doped system with K substitution on the alkaline earth site. This discrepancy  was interpreted in terms of structural modifications rather than carrier doping\,\cite{Kim2010} or by a correlation related  localization of the induced holes\,\cite{Texier2012}.

Our ARPES   study on BFCMA  reveals, contrary to the above mentioned naively expected hole doping, no decrease of the total number of charge carriers  upon Mn substitution. 
Moreover, upon Mn substitution the scattering rates and the mass renormalization do not decrease indicating that there is no pronounced coupling of spin fluctuations to the charge carriers exactly at optimal doping. 

Single crystals of BFCMA  were grown from the self-flux and characterized by resistivity, magnetization, specific heat, and dilatometry measurements\,\cite{Hardy2009}. ARPES measurements were carried out at the synchrotron radiation facility BESSY II using the UE112-PGM2a beam line and the "$1^2$"-ARPES
end station equipped with a Scienta R8000 analyzer. The total energy resolution was 10-15  meV, while the angular resolution was 0.2\dg. 
All measurements were performed in the paramagnetic and non-superconducting range near $T$=30  K. For the presentation we use a coordinate system parallel to the Fe-Fe direction as in our previous ARPES study\,\cite{Thirupathaiah2011}. 

We performed density functional theory (DFT) calculations within the generalized gradient approximation (GGA) \cite{Perdew1996} using the scalar relativistic mode of the Full Potential Local Orbitial code (FPLO) \cite{Koepernik1999}. The doping was simulated by constructing supercells with $x$=0.06 and $y$=0 or 0.07. For $k$-integration we used the tetrahedron method with $6^3$ points in the full Brillouin zone of the resulting unit cells with 16 Fe(Co/Mn) positions. We constructed 4 different cells for $y$=0.07 with different relative Co-Mn arrangements. The As-position and lattice parameters are taken from  experimental x--ray diffraction data\,\cite{Hardy2009}.

\begin{figure}
\centering
\includegraphics[angle=0,width=0.45\textwidth]{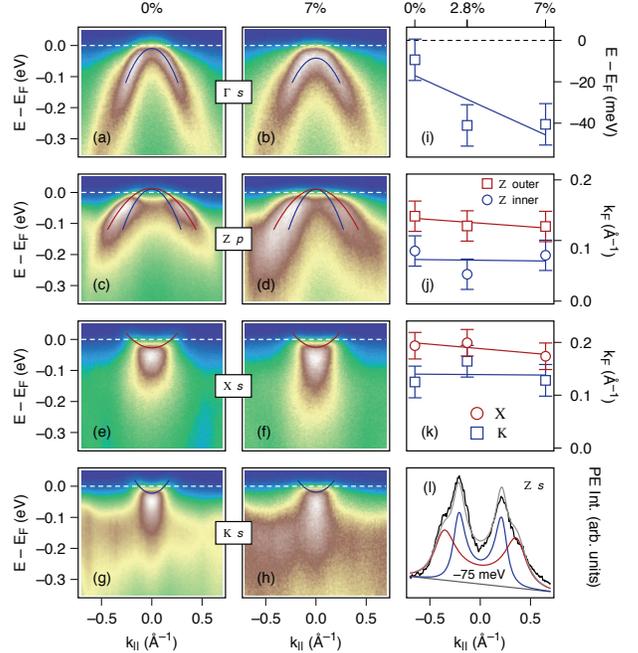}
\caption{(color online) ARPES data of \BFCMA\ . (a)-(h)  Intensity plots as a function of binding energy and wave vectors along the
$k_y$ direction  for $x=0.06$ and $y$=0 and 0.07 near the high-symmetry points $\Gamma$, $Z$, $X$, and $K$. The spectra near $\Gamma$, $X$, and $K$ were recorded with $s$-polarized photons while the spectrum near $Z$ was measured with $p$-polarized photons.  The red/blue  lines represent parabolic bands  derived from least squares fits. 
(i) Energy of the top of the inner hole bands at $\Gamma$. (j) and (k) Fermi wave vectors  near high-symmetry points as a function of Mn concentration. 
The solid lines are least-squares fits assuming a linear dependence of
$k_F$ on $y$.
(l) Momentum distribution curve at a binding energy of 75 meV  near $Z$  recorded with $s$-polaized photons.}
\label{fig1} 
\end{figure}

ARPES spectra of BFCMA $x$=0.06 and $y$= 0, 0.028, and 0.07 were measured near the four high-symmetry points $\Gamma$, $Z$, $X$, and $K$  along the $k_y$ direction using $s$ (perpendicular to $k_x$) and $p$ (parallel to $k_x$) polarized light. The high-symmetry points were reached by using variable photon energies as outlined in our previous studies of ferropnictides\,\cite{Thirupathaiah2011}. In fig.~\ref{fig1}~(a)-(h) we show representative intensity plots as a function of binding energy.
 Due to matrix element effects, in the present geometry for $s$-polarization  we detect  at $\Gamma$ and $Z$ bands which have odd  symmetry relative to the scattering plane, i.e., bands having  predominantly Fe $3d$  $xz/yz$ orbital character forming the two inner, almost degenerate hole pockets\,\cite{Thirupathaiah2011,Brouet2011}.  For $p$-polarization we detect in addition bands with predominantly $xy$ and $z^2$ (even)  orbital character, forming the outer hole pocket. At  $\Gamma$ near the Fermi level, the inner and the outer hole pockets are almost degenerate (see fig.~\ref{fig1}~(a). This was also derived  from spectra measured with $p$-polarized light (not shown). At the $Z$ point, the inner and outer hole pockets are clearly separated as seen in the intensity plots measured with $p$ polarization (see fig.~\ref{fig1}~(c)-(d) and in the momentum distribution curve measured with $s$-polarized light (see fig.~\ref{fig1}~(l)).
 At $X$ and $K$ we detect for $s$-polarization the electron pocket (see fig.~\ref{fig1}~(e)-(h)) which has at that point predominantly  $xz/yz$ character, while (not shown) for $p$--polarization the spectral weight is rather small  due to matrix element effects.

Using momentum distribution curves we have determined the dispersion of  bands, whereby we have approximated the dispersion by a second degree polynomial, The results are  shown in fig.~\ref{fig1}  by  red/blue lines. The Fermi wave vectors that can be obtained from  these dispersions are presented in fig.~\ref{fig1}~(j)-(k). Since for $x$=0.06 the inner and probably also the outer hole pockets  are  completely filled, we show for the inner hole pocket the energy of the  band maximum relative to the Fermi level (see fig.~\ref{fig1}~(i)).

\begin{figure}[tb]
\centering
\includegraphics[width=0.48\textwidth]{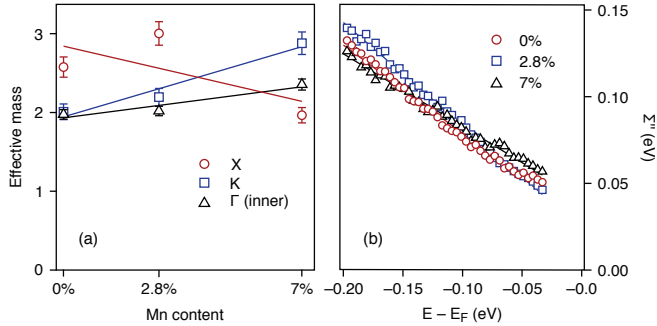}
\caption{(color online) Effective mass  and scattering rates  for BFCMA  as a function of Mn concentration. (a) effective mass $m^*/m_b$ near three high symmetry points (b)  $\Sigma$" for the inner hole pocket near the $\Gamma$ point. In both panels the solid lines are derived by  least-squares fits assuming a linear dependence.} 
\label{fig2}
\end{figure}

In fig.~\ref{fig2}~(a) we depict the effective mass $m^*/m_b$ near high symmetry points  as a function of Mn concentration.
Here $m^*$ is the mass derived from the measured dispersion and $m_b$ is the mass derived from tight-binding band structure calculations parameterized using a DFT calculation\,\cite{Korshunov2008}. In the evaluation, we used the approximation that $m_b$ is independent of the substituent concentration.
Using the measured dispersion  as an input for $\epsilon_k-\Sigma'$ ($\epsilon_k$ is the bare particle dispersion) in  the formula for the spectral function\,\cite{Huefner2003}
an MDC fit yields values for $\Sigma$" for the inner hole pocket along the $k_y$ direction near $\Gamma$ which are presented in fig.~\ref{fig2}~(b)   for three different Mn concentrations. 

First we discuss the question whether  Mn substitution into BFCMA induces holes. Already the shift of the hole pockets at $\Gamma$  to higher binding energies could signal that no holes are formed upon Mn substitution. A more detailed analysis, presented in the following,  supports this suggestion. We have evaluated the volume of the Fermi cylinders using the $k_F$ values presented in fig.~\ref{fig1}~(j)-(k). The $k_F$ values between $k_z$ = 0 and $\pi$ where interpolated by a cosine function which is a good approximation as demonstrated in our previous work\,\cite{Thirupathaiah2011} on \BFCA\ , where we have measured the full $k_z$-dispersion. Furthermore we took into account that the inner hole pocket and the electron pocket is doubly degenerate.
In this way we obtain for the optimally Co doped compound without Mn substitution a total number of electrons per transition metal site $n_e$= 0.058 $\pm$ 0.012 corresponding to an electron doping of 1.0 $\pm$ 0.2 e/Co. Using the data  from the Mn substituted samples we obtain for $y$=0.07 a decrease of the number of electrons per transition metal site of 0.009 $\pm$ 0.02 corresponding to hole doping of 0.1 $\pm$ 0.2 h/Mn. Thus from our analysis we derive that within error bars there is no hole doping upon substitution of Fe with Mn.

\begin{figure}[tb]
\centering
\includegraphics[width=0.49\textwidth]{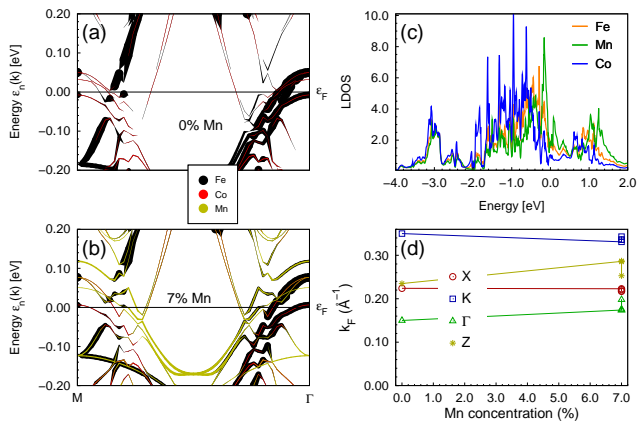}
\caption{(color online) (a,b): unfolded band structure along
$M/\Gamma$ with Co and Mn impurity weights shown in medium(red) and
light(yellow), respectively. Note that the gaps in the main Fe bands
(black) are washed out in ensemble average. (c): local density of
states (LDOS) for Co and Mn impurity bands compared to the density of states of Fe bands; (d): Fermi radii for the
outer hole and electron bands. Note, that there are 4 supercells/data-points
for $y$=0.07.} 
\label{fig3}
\end{figure}

The comparison of the ARPES data with the DFT results (see fig~\ref{fig3}) does not prove satisfactory. Contrary to the experimental data, upon Mn substitution the calculated hole pockets near $\Gamma$ move to lower binding energy (see fig.~\ref{fig3}~(a)-(b)). Moreover, also the calculated $k_F$ values (see fig.~\ref{fig3}~(d)) show an increase of the hole pockets and a small decrease of the electron pockets which would signal a considerable hole doping upon Mn substitution. A quantitative comparison between ARPES and DFT results is probably not meaningful since we know from dynamical mean field calculations\,\cite{Aichhorn2009} that correlation effects lead to a considerable renormalization of the bands which is dependent on the orbital character. On the other hand, the DFT calculations provide the important result that there is a  considerable
Mn impurity density of state near the Fermi level (see fig.~\ref{fig3}~(c)) which could explain why Mn substitution  is detrimental to superconductivity. 

Recent ARPES work\,\cite{Ideta2013} on Co, Ni, and Cu substituted \BFA\  together with theoretical work\,\cite{Wadati2010,Haverkort2011,Berlijn2012} indicated, that electron doping occurs when there is an energetic overlap of the impurity states with the Fe $3d$ states. If the impurity potential is large, i.e.,  if the impurity states are well below the Fe $3d$ band, no hybridization takes place and a localization of the additional electrons occurs. In the case of Mn substitution, one may think that the Mn impurity states could be well above Fe $3d$ band thus leading to the  localization of the holes, observed in the present study and in previous Knight shift   measurements\,\cite{Texier2012}. On the other hand, looking at  the calculated partial density of states shown in fig.~\ref{fig3}~(c), not only the Co but also the Mn impurity states overlap strongly with Fe $3d$ band states. Thus the localization of the Mn impurity holes cannot be understood on the basis of the strength of  impurity potential as in the case of Co, Ni, and Cu substitution. Probably the localization of the holes induced by Mn substitution is caused by correlation effects related to Hunds coupling, which are particularly strong for the half-filled
$3d$ shell\,\cite{Aichhorn2009,Ishida2010,Yin2011,Werner2012}.

Next we discuss the possibility of a change of  the mass renormalization as a function of the Mn concentration. We focus first on those parts of the Fermi surface which have predominantly $xz/yz$ orbital character since according to  calculations  in the random-phase-approximation~\cite{Graser2009,Hirschfeld2011} interband transitions between those bands should lead to the highest spin fluctuation susceptibilities, i.e., in particular at those hot spots quantum criticality should occur. As shown in fig.~\ref{fig2}~(a), the mass renormalization of the inner hole pocket and the electron pocket  does not show at these hot spots a decrease  which would be  expected for an increasing distance to the quantum critical point. Rather at least near  $\Gamma$ and $K$ a slight increase is realized. In addition, we emphasize that close to the Fermi level  we do not detect  a deviation from a parabolic dispersion (see fig.~\ref{fig1}~(a)) which could be related to a low--energy mass enhancement. This is a first indication that from our ARPES experiments we cannot conclude on the existence of a quantum critical point near optimal doping. One may argue that upon Mn substitution, we remain at the quantum critical point. Therefore we evaluated the effective masses of the same bands near $\Gamma$  in \BFCA\   and in \EFAP\  using our previous ARPES data\,\cite{Thirupathaiah2011}. In \BFCA\   the effective masses  remain nearly constant  when going from  optimal doping  to the overdoped case. In \EFAP\  the effective masses remain constant within 2 $\%$ when going from the undoped system (measured in the paramagnetic phase at $T$ = 220 K) to optimally substituted samples ($x$=0.16) to oversubstituted crystals ($x$=0.22). Thus, contrary to the de Haas–-van Alphen measurements on \BFAP\,\cite{Shishido2010,Shibauchi2013}, an enhanced effective mass  near optimal doping/substitution is observed in our ARPES experiments neither in \BFCA\ nor in \EFAP\  nor  in BFCMA.

We expect a corresponding result for the scattering rates. As depicted in fig.~\ref{fig2}~(b), for the inner hole pocket  $\Sigma$"=$\alpha+\beta E$ where $\alpha \approx$ 40 meV  is determined by elastic scattering due to defects, e.g. by a disordered Ba surface\,\cite{Heumen2011}. The coefficient $\beta \approx$ 0.5 is a measure for the inelastic scattering rate related to electron-electron interaction, e.g. via a coupling to excitations of spin fluctuations. Our data on $\Sigma$" reveal the same energy dependence for all Mn concentrations. It is remarkable that an evaluation of $\Sigma$" of our previous \BFCA\  data for $x$=0.06 to 0.2   yield a 20 \% \emph{increase} instead of  a decrease of the  $\beta$ values, when going from optimally doped to overdoped samples. Thus also the scattering rates derived from our ARPES experiments  on this system do not diverge at optimal doping and thus do not support a quantum critical point where a particularly large  coupling of  spin fluctuations   to the charge carriers is expected.  

One  may argue that in the present ARPES experiments the temperature is too high to detect indications of quantum criticality in the pnictides. We emphasize that the strange behavior of the thermodynamic and transport properties was detected at comparable  temperatures where the present  information on the scattering rates was obtained.

 Since the thermodynamic and the transport properties may be determined by the cold spots of the Fermi surfaces we also discuss the renormalization of these bands although there is less information available from our ARPES measurements. The outer hole pocket has predominantly $xy$ orbital character and according to theoretical results~\cite{Graser2009,Hirschfeld2011} the susceptibility for spin fluctuations should be considerably smaller for this band. Indeed our ARPES data show strongly reduced scattering rates for the outer hole pocket although it is difficult to evaluate exact values for $\beta$. Finally, according to the RPA calculations~\cite{Graser2009,Hirschfeld2011}, along the $k_x$ direction there should be also a part of the electron pocket with predominantly $xy$ orbital character which should have a lower renormalization caused by  electron-electron interaction. This is indeed observed in ARPES experiments by Brouet et al.\,\cite{Brouet2011} where the scattering rate  of the electron pocket along $k_y$ ($yz$ character) is about twice as large as that along the $k_x$ direction ($xy$ character). Thus measured scattering rates do not indicate a strong renormalization for the cold spot with $xy$ character at optimal doping.

In the present ARPES study of  BFCMA  at the hot spots we do not detect a particularly strong coupling to spin fluctuations at optimal doping which is expected  in a quantum critical scenarios for unconventional superconductivity. The reason for this is probably that the energy scale of the spin fluctuation excitations, coupled to the charge carriers in a dynamical nesting process is rather high compared to the energetic changes of the band structure caused by Fe substitution. This view is supported by very recent resonant inelastic x-ray scattering study in which spin excitations in the energy range 150-200 meV were detected with minor changes when going from the undoped AFM compound to the paramagnetic superconducting system\,\cite{Zhou2013}. In this way it is possible to understand the continuous development of the renormalization of the electronic structure in a large range of the control parameter  which was also observed for the Drude  scattering rates derived from optical spectroscopy\,\cite{Nakajima2010}.

In this context, it is interesting that in other systems the strange properties near a quantum critical point are often discussed  in terms of a single electron model, e.g. by a band edge or a singularity in the density of states and/or changes in the topology of the Fermi surface\,\cite{Pfau2012}. In the ferropnictides such scenarios are possible since near optimal doping a  Lifshitz transition is observed: the Fermi surface changes from a cylinder along the $k_z$ direction to an ellipsoid around the $Z$ point\,\cite{Liu2010,Thirupathaiah2011} (see also fig.~\ref{fig1}~(a)-(b)). This observation may indicate the importance of this Lifshitz transition for the magnetic and the superconducting properties of pnictides.

\acknowledgments
We acknowledge funding by the DFG through the priority program SPP1458. We thank Peter Schweiss for providing x-ray diffraction data of the samples and Roman Schuster for helpful discussions.

\bibliographystyle{epl}

\end{document}